\documentclass[conference]{IEEEtran}

\IEEEoverridecommandlockouts
\usepackage{cite}
\usepackage{amsmath,amssymb,amsfonts}
\usepackage{graphicx}
\usepackage{textcomp}
\usepackage{xcolor}
\usepackage{booktabs}
\usepackage{url}
\usepackage{array}
\usepackage{xspace}

\def\BibTeX{{\rm B\kern-.05em{\sc i\kern-.025em b}\kern-.08em
    T\kern-.1667em\lower.7ex\hbox{E}\kern-.125emX}}
\newcommand{\qev}{QEVOLVE-Bench\xspace}
\newcommand{\agent}{QEVOLVE-Agent\xspace}
\begin{document}
\title{QEVOLVE-Bench: A Seed Benchmark for Quantum SDK Evolution and Repair Planning}
\author{
\IEEEauthorblockN{Krishna Bhatia}
\IEEEauthorblockA{\textit{QuantumAI Lab, Fractal Analytics}\\
India\\
krishna.bhatia@fractal.ai}
\and
\IEEEauthorblockN{Gautami Sanjay Naik}
\IEEEauthorblockA{\textit{Indian Institute of Information Technology Dharwad}\\
India\\
gautaminaik2000@gmail.com}
}
\maketitle
\begin{abstract}
Quantum software is increasingly built on fast-moving Python SDKs such as Qiskit, PennyLane, and Cirq. When these SDKs evolve, user programs can fail because execution helpers are removed, import paths change, simulator abstractions shift, device names are deprecated, or circuit export interfaces are revised. Although some failures surface as simple missing-symbol errors, repairing them is not only a syntactic problem: developers must identify the replacement workflow, preserve the quantum-facing intent, and validate behavior under pinned framework versions. This paper presents \qev, a small executable seed benchmark for quantum SDK evolution. The current public release contains 14 controlled migration tasks across Qiskit, PennyLane, and Cirq. Each task is a self-contained Python project with pinned dependencies, failing and passing outputs, pytest acceptance checks, metadata, and benchmark notes. The artifact also includes a manifest, sanity checker, representative smoke-test logs, Zenodo archive, screencast, and a lightweight \agent prototype that converts task evidence into structured repair proposals. \qev is intentionally not a statistically powered benchmark; instead, it is an extensible seed artifact that lowers the setup cost for studying automated repair, LLM-based software engineering, and maintenance of quantum SDK-based projects.
\end{abstract}


\section{Introduction}
Quantum programming frameworks are moving targets. Qiskit, PennyLane, and Cirq provide the abstractions used to construct circuits, select simulators or devices, export programs, and execute experiments. They also evolve as the quantum-software stack matures. For developers, this creates a recurring maintenance problem: a program can fail not because the underlying quantum algorithm is wrong, but because the SDK removed an execution helper, renamed an import path, changed a simulator abstraction, deprecated a device name, or revised a circuit-export interface.

This paper focuses on that maintenance setting: \emph{quantum SDK evolution}. At first glance, API-evolution failures may appear trivial because a static checker or interpreter can often detect a missing function or import. Detection, however, is only the first step. A repair tool must still determine the replacement idiom for a particular SDK version, update the surrounding workflow, and validate that the circuit, measurement, export, or simulator behavior remains acceptable. For example, replacing a removed call to a high-level execution helper may require moving to a backend/transpile/run workflow; replacing a circuit-export method may require selecting the correct OpenQASM interface; and replacing a deprecated device name may require checking that the test still validates the intended quantum behavior. These are small tasks, but they are domain-shaped tasks.

General software-engineering benchmarks such as Defects4J~\cite{just2014defects4j}, QuixBugs~\cite{lin2017quixbugs}, and SWE-bench~\cite{jimenez2024swebench} show the value of executable tasks for testing and repair research. Quantum software engineering has also developed artifacts and studies around platform bugs, quantum-program bugs, testing, and analysis~\cite{paltenghi2022bugs,zhao2023bugs4q,paltenghi2024survey,destefano2022quantumprogramming}. Yet there is still a gap for a small, reusable artifact focused specifically on user-code maintenance under changing quantum SDK interfaces.

We present \qev, an executable seed benchmark for this gap. The current release contains 14 controlled SDK-migration tasks across Qiskit, PennyLane, and Cirq. The goal is not to claim broad statistical coverage or state-of-the-art repair results. Rather, \qev packages a reproducible task format, public artifact, and prototype repair-planning interface that can be extended into a larger benchmark. The artifact is available on GitHub and archived on Zenodo, with a screencast demonstrating its structure and use.

The contributions are: (1) an executable seed benchmark for quantum SDK evolution tasks; (2) a task schema and repository layout with pinned requirements, failing/passing evidence, and acceptance tests; and (3) a lightweight \agent interface that converts task metadata and failure evidence into structured repair proposals for future repair workflows.

\section{Benchmark Scope and Task Design}
\qev is organized as a repository containing controlled seed projects, documentation, scripts, and outputs. The main inventory is a CSV manifest that records task identifier, ecosystem, migration category, task directory, and validation status. Each task is independently executable and uses the same minimal structure:

\begin{table}[t]
\caption{Per-task structure in \qev.}
\label{tab:schema}
\centering
\footnotesize
\begin{tabular}{ll}
\toprule
File or directory & Purpose \\
\midrule
\texttt{task.yaml} & benchmark metadata \\
\texttt{requirements.txt} & pinned task dependencies \\
\texttt{src/} & source code under repair \\
\texttt{tests/} & pytest acceptance checks \\
\texttt{failing\_output.txt} & captured pre-repair failure \\
\texttt{passing\_output.txt} & captured post-repair success \\
\texttt{benchmark\_notes.md} & failure signal and intended fix \\
\bottomrule
\end{tabular}
\end{table}

\begin{table}[t]
\caption{Current public task inventory.}
\label{tab:inventory}
\centering
\footnotesize
\begin{tabular}{lrl}
\toprule
Ecosystem & Count & Example migration families \\
\midrule
Qiskit & 7 & execution, QuantumInstance, QASM export, imports \\
PennyLane & 5 & legacy devices, execute keywords, operator helpers \\
Cirq & 2 & deprecated base classes and transformer helpers \\
\midrule
Total & 14 & controlled SDK-evolution tasks \\
\bottomrule
\end{tabular}
\end{table}

Table~\ref{tab:schema} shows the common task structure, and Table~\ref{tab:inventory} summarizes the current release. The tasks are deliberately small and simulator-only. They do not require hardware credentials, cloud backends, or long-running quantum simulations. This makes them suitable for workshops, tutorials, repair-tool debugging, and repeated execution inside automated agents.

The current categories were selected to represent maintenance-level SDK evolution rather than algorithmic redesign. Qiskit tasks include execution workflow changes, parameter binding migration, OpenQASM export migration, and import-path drift. PennyLane tasks include legacy device-name changes, execution keyword migration, and operator-helper changes. Cirq tasks include deprecated base-class and helper-protocol migrations. The benchmark therefore captures cases where the quantum intent is stable but the SDK-facing implementation must evolve.

\section{Why This Lowers Evaluation Effort}
A small seed benchmark is useful only if it reduces work for future tool builders. \qev does this in four ways. First, it removes task-discovery overhead. A researcher does not need to search through SDK changelogs and construct failing projects from scratch; the manifest exposes concrete migration tasks with categories and directories. Second, it removes environment guesswork. Each task includes pinned requirements and captured failing/passing outputs. Third, it provides executable acceptance checks rather than prose-only bug descriptions. Fourth, it supplies structured metadata and notes that can be consumed by repair tools.

This design also clarifies what static analysis can and cannot do. Static or dynamic checks can often expose a missing symbol, but they do not define the correct replacement workflow, the expected test oracle, or the framework version in which the behavior should be validated. \qev packages those elements together. The benchmark is therefore not mainly a detector benchmark; it is a repair-and-validation substrate for quantum SDK evolution.

A user can begin with the repository-level sanity checker:

\begin{center}
\footnotesize
\texttt{python scripts\textbackslash smoke\_check.py}
\end{center}

The expected output reports 14 tasks and the ecosystem distribution \{Qiskit: 7, PennyLane: 5, Cirq: 2\}. To execute an individual task, the user enters the seed-project directory, creates a virtual environment, installs the pinned requirements, and runs pytest. Representative smoke logs are included for one task from each supported ecosystem. These logs are not a full empirical evaluation; they are artifact-quality evidence that the task format is executable across Qiskit, PennyLane, and Cirq.

\begin{figure}[t]
  \centering
  \includegraphics[width=0.98\linewidth]{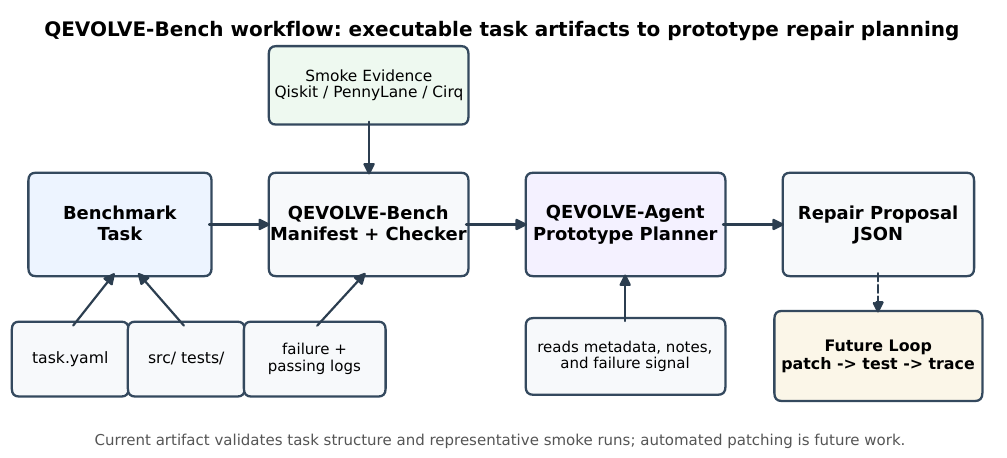}
  \caption{\qev workflow. Executable task folders expose metadata, source/tests, and failure evidence. The manifest and checker validate the artifact structure. The prototype planner consumes task evidence and emits structured repair proposals; automated patching and iterative test feedback are future extensions.}
  \label{fig:workflow}
\end{figure}

\section{Prototype Repair-Planning Interface}
The artifact includes a lightweight prototype, \agent, implemented as a repository script. The prototype reads either a task directory or the benchmark manifest, extracts metadata, benchmark notes, and failing output, and writes a JSON repair proposal. Each proposal records the task id, ecosystem, category, observed failure signal, generic repair steps, category-specific hints, and evidence files consulted.

Figure~\ref{fig:workflow} summarizes this artifact-to-planner workflow. The current prototype is intentionally modest. It does not apply patches and does not claim autonomous repair performance. Its purpose is to show that \qev tasks expose enough structure for future rule-based, LLM-based, or agentic repair systems: load metadata, inspect failure evidence, infer a migration category, propose a repair plan, apply a patch, run tests, and record a trace.

This positioning follows the broader movement toward execution-backed repair evaluation. SWE-bench evaluates language models on real GitHub issue tasks~\cite{jimenez2024swebench}; SWE-agent studies interfaces for language-model software-engineering agents~\cite{yang2024sweagent}; and RepairAgent demonstrates autonomous LLM-based program repair with tool use and test feedback~\cite{bouzenia2025repairagent}. \qev brings a similar task-consumption style to a quantum-SDK evolution setting, while separating the benchmark artifact from any claim of mature autonomous repair.

\section{Evaluation Scenarios Enabled by the Artifact}
Although this release does not report a comparative repair study, its structure is intended to make such a study straightforward. A future evaluator can run the same task set against rule-based migrations, retrieval-augmented LLM prompts, or agentic repair loops. For each task, the evaluator can measure whether the final test suite passes, how many patch attempts were required, which evidence files were used, and whether the proposed repair matches the documented migration family. This makes the artifact useful even before it reaches the scale of large software-engineering benchmarks.

\begin{table}[t]
\caption{Example benchmark tasks and the repair knowledge they exercise.}
\label{tab:examples}
\centering
\footnotesize
\begin{tabular}{p{0.20\linewidth}p{0.26\linewidth}p{0.42\linewidth}}
\toprule
Task family & Surface failure & Required repair knowledge \\
\midrule
Qiskit QASM export & removed or changed export method & select the version-appropriate OpenQASM export interface and preserve circuit text checks \\
Qiskit execution & missing execution helper & migrate from a removed convenience API to backend, transpilation, run, and result handling \\
PennyLane device & deprecated legacy device name & choose the supported device while preserving QNode behavior and acceptance checks \\
Cirq helper API & renamed/deprecated helper & update the helper call and validate circuit/state behavior under the new interface \\
\bottomrule
\end{tabular}
\end{table}

Table~\ref{tab:examples} illustrates why these tasks are not meant to stop at error detection. The missing symbol is often obvious, but the benchmark asks whether a repair tool can identify and validate the correct SDK-specific replacement. This distinction matters for LLM and agentic tools: a plausible textual patch is insufficient unless the pinned task environment and acceptance checks confirm it.

The included \agent output schema also supports trace-based evaluation. A repair system can store the observed failure, retrieved notes, proposed migration, patch, test result, and final status. Such traces would make it possible to compare repair workflows without changing the benchmark task format. In this sense, \qev is a seed for a larger evaluation framework rather than only a collection of examples.

\section{Positioning Against Related Artifacts}
\qev is closest in spirit to executable bug and repair benchmarks, but differs in domain and failure mode. Defects4J provides real Java bugs for reproducible testing studies~\cite{just2014defects4j}; QuixBugs provides small program-repair tasks~\cite{lin2017quixbugs}; and SWE-bench provides issue-driven repository tasks for evaluating language models~\cite{jimenez2024swebench}. \qev adopts the reproducibility principle of these benchmarks but targets quantum SDK evolution rather than general-purpose defects.

Within quantum software engineering, Bugs4Q provides real quantum-program bugs with tests~\cite{zhao2023bugs4q}, while platform-bug studies characterize defects inside quantum computing frameworks~\cite{paltenghi2022bugs}. Surveys and reviews show that quantum software engineering is still developing testing, debugging, and maintenance methods~\cite{paltenghi2024survey,destefano2022quantumprogramming,mandal2025review}. \qev is complementary: it focuses on user-code maintenance caused by SDK/interface drift, not platform-internal bugs or quantum-algorithm defects.

\qev also differs from quantum SDK and language artifacts. Qiskit, PennyLane, Cirq, and OpenQASM provide programming abstractions and execution/export interfaces~\cite{javadiabhari2024qiskit,bergholm2018pennylane,cirq2026,cross2022openqasm}. Benchpress evaluates performance and functionality across quantum software development kits~\cite{nation2025benchpress}. These works define or measure the software stack itself. \qev instead captures small cases where user code must evolve as those stacks change.

\section{Limitations, Availability, and Roadmap}
The current release has clear limitations. It contains only 14 controlled tasks, including only two Cirq tasks. It should therefore be treated as a seed benchmark, not as a statistically powered evaluation suite. The tasks are small and do not yet include real repository histories. Passing a task means satisfying the included executable checks, not proving full semantic equivalence for arbitrary quantum programs. The current \agent prototype is a repair planner, not a validated autonomous repair system.

The public artifact is available at \url{https://github.com/FraQTech/qevolve} and archived at \url{https://doi.org/10.5281/zenodo.20134741}. A screencast is available at \url{https://youtu.be/2KIo3T60Lc8}. The repository contains the task suite, manifest, per-task requirements, smoke-test logs, documentation, and prototype repair-planning script.

The roadmap is to expand the benchmark with more tasks, add real repository cases, increase Cirq coverage, package the execution environment more fully, integrate repair baselines, and record structured patch/test traces. In future releases, stronger semantic checks such as unitary or state-vector equivalence can complement pytest-level acceptance checks where appropriate.

\section{Conclusion}
\qev is an executable seed benchmark for quantum SDK evolution and repair planning. It packages 14 controlled migration tasks across Qiskit, PennyLane, and Cirq with pinned dependencies, failure evidence, passing evidence, metadata, tests, documentation, smoke logs, and a prototype repair-planning interface. The contribution is intentionally scoped: \qev is not yet a mature large-scale benchmark, but it is a public and archived starting point that lowers the cost of developing and evaluating repair workflows for quantum SDK maintenance.

\bibliographystyle{IEEEtran}
\bibliography{references}

\end{document}